\documentclass[preprint,12pt,authoryear]{elsarticle}
\usepackage[top=3cm,bottom=2.8cm,left=3cm,right=3cm]{geometry}

\usepackage{amssymb}
\usepackage{url}

\usepackage{tikz}

\newcommand*\circled[1]{\smash{\tikz[baseline=(char.base)]{
            \node[shape=circle,draw,inner sep=0pt,text=black,minimum size=1.1em] (char) {#1\strut};}}}

\usepackage[most]{tcolorbox}
\newtcolorbox[auto counter]{info_box}[2][]{title=Box~\thetcbcounter: #2, #1}

\usepackage{caption}
\DeclareCaptionLabelFormat{one-page}{\hrulefill\\#1 #2}
\DeclareCaptionLabelFormat{two-page}{\hrulefill\\#1 #2 \emph{(previous page)}}

\usepackage{natbib} 

\usepackage{siunitx}

\begin{document}

\begin{frontmatter}



\title{The era of the ARG: an empiricist's guide to ancestral recombination graphs}


\author[INT_MSU,KBS_MSU,EEB_MSU,EEB_UMICH]{Alexander L. Lewanski}
\ead{allewanski@gmail.com}
\author[EEB_UMICH]{Michael C. Grundler}
\author[KBS_MSU,EEB_UMICH]{Gideon S. Bradburd}

\affiliation[INT_MSU]{organization={Department of Integrative Biology, Michigan State University},
            city={East Lansing},
            state={MI},
            country={US}}

\affiliation[KBS_MSU]{organization={W.K. Kellogg Biological Station, Michigan State University},
            city={Hickory Corners},
            state={MI},
            country={US}}

\affiliation[EEB_MSU]{organization={Ecology, Evolution, and Behavior Program, Michigan State University},
            city={East Lansing},
            state={MI},
            country={US}}

\affiliation[EEB_UMICH]{organization={Department of Ecology and Evolutionary Biology, University of Michigan},
            city={Ann Arbor},
            state={MI},
            country={US}}

\begin{abstract}
In the presence of recombination, the evolutionary relationships between a set of sampled genomes cannot be described by a single genealogical tree. Instead, the genomes are related by a complex, interwoven collection of genealogies formalized in a structure called an \textit{ancestral recombination graph} (ARG). An ARG extensively encodes the ancestry of the genome(s) and thus is replete with valuable information for addressing diverse questions in evolutionary biology. Despite its potential utility, technological and methodological limitations, along with a lack of approachable literature, have severely restricted awareness and application of ARGs in empirical evolution research. Excitingly, recent progress in ARG reconstruction and simulation have made ARG-based approaches feasible for many questions and systems. In this review, we provide an accessible introduction and exploration of ARGs, survey recent methodological breakthroughs, and describe the potential for ARGs to further existing goals and open avenues of inquiry that were previously inaccessible in evolutionary genomics. Through this discussion, we aim to more widely disseminate the promise of ARGs in evolutionary genomics and encourage the broader development and adoption of ARG-based inference.




\end{abstract}




\end{frontmatter}



\section{Introduction}
\label{sec:intro}
Many of the principal pursuits in evolutionary genomics can be recast as questions about the transmission of genetic material from ancestors to descendants. For example, in the study of speciation and hybridization, we may be interested in identifying which sections of a hybrid genome were derived from which parental species \citep{Marques2019,Moran2021TheHybridization}. As another example, we often want to know about the nature of selection on a genetic variant \citep[e.g.,][]{Martinez-Jimenez2020AGenes,Schluter2022ThreeSelection,Henn2015EstimatingGenomes,Barrett2008AdaptationVariation}, which is, in essence, asking whether the variant has displayed a particular pattern of transmission. For instance, a positively selected variant confers a fitness advantage and thus would be preferentially transmitted between generations. In applied settings, we may want to understand whether a human-made structure such as a road or dam \citep[e.g.,][]{Epps2005HighwaysSheep,Machado2022DammingPopulations} reduces connectivity between populations, which is implicitly asking how often ancestor-descendant relationships span the potential barrier \citep[e.g.,][]{Jasper2019AMosquito}. So far, direct knowledge of how genetic material is transmitted from ancestors to descendants is extremely limited in nearly all systems, save those with extensive pedigree and genomic information [e.g., Florida Scrub-jays \citep{Chen2016,Aguillon2017DeconstructingDispersal,Chen2019AllelePopulation}, economically important livestock like dairy cattle \citep{Larkin2012Whole-genomeCattle,Ma2015CattleAnalysis}]. However, access to this information could revolutionize the study of numerous topics across evolutionary genomics.

In population genetics, the central structure that describes how genetic material is passed from ancestors to descendants is called an \textit{ancestral recombination graph} (ARG). Building on earlier developments in coalescent theory \citep{Kingman1982TheCoalescent, Kingman1982OnPopulations,Tajima1983EvolutionaryPopulations,Hudson1983PropertiesRecombination}, ARGs were conceptualized in the 1990s by R.C. Griffiths and P. Marjoram \citep{Griffiths1991TheGraph, Griffiths1996AncestralRecombination, Griffiths1997AnGraph} to describe ancestry in the presence of coalescence and recombination. ARGs have subsequently featured prominently in the theoretical and statistical realms of population genetics where they have been extensively studied for their biological, mathematical, and computational properties and utility.

In contrast, ARGs remain much less known and appreciated in empirical evolutionary genomics. This inattention can at least partially be ascribed to pragmatism---until recently, ARGs have been purely theoretical constructs, impractical to reconstruct in empirical systems or even simulate at biologically realistic scales. Additionally, although an expansive literature already exists on ARGs, much of this content is targeted at an audience with an extensive theoretical or statistical background in population genetics and thus may be unapproachable for some empirical biologists. Excitingly, recent methodological advances in reconstructing (Box \ref{box_arg_reconstruction}) and simulating (Box \ref{box_arg_simulation}) ARGs together with concurrent progress in genome sequencing and increasingly available high-performance computation means that ARG-based inference is rapidly becoming attainable in empirical- and simulation-based evolutionary genomics research. To help usher in this imminent ``era of the ARG,'' we view now as an opportune moment to provide a widely accessible resource for comprehending ARGs and their potential in evolutionary genomics.

We have two primary objectives for this paper. First, we provide a concise and gentle primer on ARGs, including an introduction to what an ARG is, what information can be encoded within it, and an exploration of some of its basic properties. Second, we discuss the current and future potential for ARGs to benefit evolutionary genomics research. Our aim for the second objective is not to exhaustively review existing ARG-based research, but rather to articulate the promise of ARGs to advance diverse topics across evolutionary genomics. We supplement these two main objectives with an overview of recent methodological developments in inferring, simulating, and analyzing ARGs. This discussion will demonstrate the current or impending feasibility of ARG-based inference for many evolutionary genomics questions and systems.

\section{An ARG primer}
\label{sec:arg_primer}
In the following section, we will incrementally develop an intuition for what ARGs are by starting with the fundamentals of sexual reproduction and genealogical relatedness, which will help clarify how ARGs emerge from these first principles of biology. To simplify our discussion, we will focus on the nuclear genome of sexual, diploid organisms and meiotic recombination throughout the paper. However, the ideas covered here are relevant to any organism across the tree of life as well as viruses whose genomes undergo any type of recombination (e.g., gene conversion, bacterial conjugation). For more technical treatments of ARGs, we direct interested readers to \cite{Griffiths1997AnGraph}, \cite{Wiuf1999RecombinationSequences}, \cite{Hein2005TheRecombination}, and Wong et al. (unpublished). 

\subsection{Background}
In sexual, diploid organisms, haploid gametes are generated by the sampling of a single DNA copy of every position in the genome during meiosis. During reproduction, the parents’ gametes fuse, which leads to a diploid offspring. The relationships between a set of individuals can be represented by a genealogical pedigree (Figure \ref{arg_review_main_figure_genomearg}A), in which each individual has two parents, from each of whom it has inherited exactly half of its genome. The pedigree consists of nodes, which represent individual organisms, and edges, which connect a subset of the nodes and signify parent-offspring relationships.

\begin{figure}[btp]
    \captionsetup{labelformat=one-page}
    \centering
    \includegraphics[width=\textwidth]{figures/pdfs/arg_review_main_figure_genomearg_V2.pdf}
    \caption{} 
\end{figure}
\begin{figure}
    \captionsetup{labelformat=two-page}
    \ContinuedFloat
    \caption{Overview of \textit{ancestral recombination graphs} (ARG). In all ARG depictions (A, B, D), nodes are indicated by small circles and each represents a single set of one or more chromosomes (a haploid genome) of an individual. The node coloration indicates whether or not it is involved in recombination, and the specific pattern (shading and outline) of the node indicates its type: nonsample, unary (nonsample), sample. The genome is divided into three non-recombining regions (blue, orange, and green). (A) The relationships of multiple individuals can be organized into a pedigree. An ARG is embedded in a pedigree and represents the set of pedigree paths through which genetic material is transmitted. (B) The graphical representation of an ARG. Edges (the connections between nodes) are colored and annotated with the non-recombining region(s) that they transmit. (C) A plot recording the lineage count through time in the ARG. Backward in time, coalescent events, which occur at the dark gray points, merge lineages and thus reduce the lineage count. The red points highlight the times at which recombination occurs, which splits lineages backward in time and therefore increases the lineage count. (D) An ARG can be formulated as a series of local trees that share nodes and edges. Each non-recombining region possesses its own local tree. The regions are separated by a recombination event, which, when moving between regions, prunes a portion of the tree and regrafts it to another node. This action means that nearby trees are generally quite similar in structure. The arrows in the left two trees show how recombination relocates a branch in the tree (reconnecting to the small, light gray node) to form the tree of the region immediately to the right. The dashed lines on the second and third trees highlight each tree's shared structure with its leftward neighbor.}
    \label{arg_review_main_figure_genomearg}
\end{figure}

By itself, the pedigree can provide coarse estimates of genetic ancestry, such as the expected genetic relatedness between individuals (e.g., 0.50 between full siblings; 0.125 between first cousins), or the expected proportion of the genome inherited from a particular genealogical ancestor. However, for any region of the genome, we are unable to ascertain from the pedigree alone whether it is the parent's maternal or paternal copy that has been transmitted. Thus, we are restricted to calculating expected quantities. We could therefore gain more in-depth knowledge of ancestry in the genome by explicitly tracking the transmission of DNA sequences down the pedigree from specific parental to offspring chromosomes.

This discussion of the pedigree highlights multiple key ideas in our build-up to ARGs. First, because each parent contributes only one DNA copy at a particular genomic position to its offspring, each copy (including copies contained within an individual) experiences its own unique history of inheritance through the pedigree. Second, because a parent only contributes half of its genome to each offspring and not all individuals reproduce, only a subset of the genetic material possessed by historical individuals in the pedigree end up in contemporary individuals. As you travel further back in the pedigree, despite the geometric increase in the number of expected genealogical ancestors [$2^n$ ancestors (assuming no inbreeding) where $n$ equals the number of generations back in time], an increasing proportion of these ancestors contributes no genetic material to their contemporary descendants \citep{Donnelly1983TheDescent,Chang1999RecentIndividuals}. 

If we concentrate on a particular position in an individual's genome, we see that each DNA copy traverses just one of the manifold possible paths (i.e., series of connected nodes and edges) in the pedigree. The specific pedigree paths through which copies at a particular position in contemporary individuals were transmitted from their ancestors represent the genetic genealogy at that position \citep{Hudson1991GeneProcess,Mathieson2020}. Similar to a pedigree, each edge in the genealogy represents a transmission event of genetic material from parent to offspring. However, in a pedigree, each node is a diploid individual, while in a genetic genealogy, each node represents one of two haploid sequences \textit{within} a diploid individual---the specific genomic copy sampled to create a gamete that passes genetic material from a parent to the current individual. This genetic genealogy is embedded in the pedigree (Figure \ref{arg_review_main_figure_genomearg}A). The sequence of relationships defined by the pedigree constrains the possible nodes and edges that can exist in the genealogy, but does not fully dictate the identity of these nodes and edges. The structure of a genetic genealogy is determined by both the pedigree structure and the outcome of the gametogenic genome sampling at each reproduction event in the pedigree.

The genetic perspective of relatedness is further complicated by another feature of meiosis: recombination. Meiotic recombination, the shuffling of genetic material in the genome during meiosis, occurs via two processes: (1) exchange of genetic material between homologous chromosomes via crossing over during prophase I; (2) random assortment of homologous chromosomes during anaphase I. These recombinational processes can produce a mosaic of genetic ancestry across the haploid genome of the gamete so that a particular gametic genome potentially contains genetic material inherited from different parents both between non-homologous chromosomes and within chromosomes. Recombination therefore results in different histories of inheritance (and thus different genealogies) across the genome, with topological changes to the genealogy associated with recombination breakpoints and different chromosomes \citep{Rosenberg2002GenealogicalPolymorphisms}.

\subsection{Ancestral recombination graphs}
The complex web of genetic genealogies across the genome is recorded in a graphical structure known as an \textit{ancestral recombination graph} (ARG), which provides extensive information regarding the history of inheritance for a set of sampled genomes. Each node in an ARG represents a haploid genome (a \textit{haplotype}) in a real individual that exists now or in the past (Wong et al., unpublished). Each diploid individual therefore contains two haploid genomes and is represented by two nodes. We refer to nodes corresponding to sampled genomes [often, though not necessarily \citep[e.g.,][]{Schaefer2021AnGenomes,Speidel2021InferringGenealogies,Wohns2022AGenomes}, sampled in the present] as \textit{sample nodes} and all other nodes as \textit{nonsample nodes}. If sample nodes have no sampled descendants, they constitute the tips of an ARG. Sample nodes are particularly salient because ARGs are generally specified in terms of the genetic ancestry of these genomes. Edges in an ARG indicate paths of inheritance between nodes. ARGs are technically described as ``directed graphs'' because genetic material flows unidirectionally from ancestors to descendants. 

Assuming that sample nodes are sourced from contemporary individuals, the present time in an ARG (the bottom of the vertical axes in Figures \ref{arg_review_main_figure_genomearg}B and D) contains a lineage (i.e., sets of one or more edges connected by nodes forming continuous paths of inheritance) for each sample. Tracing the lineages back in time, some nodes have two edges enter on the future-facing side but only a single outbound edge on the past-facing side (e.g., node \circled{\footnotesize R} in Figure \ref{arg_review_main_figure_genomearg}B). These nodes represent haplotypes in which two lineages find common ancestry and thus merge into a single lineage, which reduces the lineage count by one (the dark gray points in Figure \ref{arg_review_main_figure_genomearg}C). Common ancestry events additionally represent \textit{coalescence} when (backward in time) the two merging edges contain the same portion of the genome [note that all nodes corresponding to common ancestry events in Figure \ref{arg_review_main_figure_genomearg} (\circled{\footnotesize K}, \circled{\footnotesize P}, \circled{\footnotesize Q}, \circled{\footnotesize R}, \circled{\footnotesize W}, and \circled{\footnotesize X}) also correspond to coalescence]. From an organismal perspective, nodes corresponding to coalesence represent an instance in which a parent provides the same (portion of a) haploid genome to multiple offspring and thus splits a lineage into multiple lineages forward in time.

Conversely, other nodes have a single edge enter on the future-facing side but two edges exit the past-facing side (e.g., node \circled{\footnotesize Q} in Figure \ref{arg_review_main_figure_genomearg}B), which represents the outcome of recombination \citep{Griffiths1997AnGraph}. Backward in time, the node with two outbound edges on the past-facing side is the recombinant offspring node whose genome is inherited from two parental nodes (e.g., node \circled{\footnotesize C} in Figure \ref{arg_review_main_figure_genomearg}). The two nodes that each receive one of the outbound edges are the parental nodes whose genomes are recombined in the offspring node. For example, in Figure \ref{arg_review_main_figure_genomearg}, \circled{\footnotesize G} and \circled{\footnotesize H} are the parental nodes of \circled{\footnotesize C}. From an organismal perspective, these nodes occur when an offspring receives one of its haploid genomes from a parent, and that haploid genome represents the outcome of recombination between the parent's two haploid genomes. Recombination splits the genome into separate lineages and thus each portion of the genome experiences a distinct history of inheritance between (traversing an ARG from present to past) the recombination event from which they split to the coalescence event in which they join back up. Consequently, each recombination event increases the number of lineages in an ARG by one (\citealt{Nordborg2001CoalescentTheory}; the red points in Figure \ref{arg_review_main_figure_genomearg}C). From a forward-in-time perspective, recombination fuses portions of two parental genomes into a single haplotype (in the recombinant offspring), and thus unites separate lineages into a single lineage. Nodes through which ancestral material is transmitted but are involved in neither coalescence nor recombination for genomic material that is ancestral to the samples do not determine the topology of an ARG and thus are frequently omitted (we retain several of these nodes in Figure \ref{arg_review_main_figure_genomearg} to highlight the effects of recombination). More generally, nodes with only one descendant (\textit{unary} nodes; e.g., node \circled{\footnotesize S} in Figure \ref{arg_review_main_figure_genomearg}) do not directly influence genealogical relationships between the sample nodes. In simulations, unary nodes are often removed via a process called \textit{simplification} \citep{Kelleher2018EfficientSimulation}, and in empirical ARGs, these are not even inferred. 

ARGs record the timing of each node and the portion of the genome that each edge transmits between ancestors and descendants. To trace the genealogy for a particular position in the genome, you follow the edges through the ARG that contain the focal position \citep{Griffiths1997AnGraph}. For example, in Figure \ref{arg_review_main_figure_genomearg}B, if you want to extract the genealogy for a position in the orange region (between positions 0 and 1) of sample node \circled{\footnotesize B}, you would follow the edges that transmit the orange region between nodes (i.e., \circled{\footnotesize B} $\rightarrow$ \circled{\footnotesize K} $\rightarrow$ \circled{\footnotesize R} $\rightarrow$ \circled{\footnotesize W} $\rightarrow$ \circled{\footnotesize X}). If the entire genome finds common ancestry, the first common ancestor is called the \textit{most recent common ancestor} (MRCA) of the genome or the \textit{Grand MRCA} (GMRCA; \citealt{Griffiths1997AnGraph}). 

The fact that each genomic region bracketed by recombination breakpoints (hereafter \textit{non-recombining region}) possesses its own genealogy and that a non-recombining region in a single sample node traces only one path back to the MRCA of the entire sample suggests an alternative representation of an ARG: an ordered set of genealogical trees along the genome with labeled sample and nonsample nodes to specify how nodes are shared between trees (\citealt{Griffiths1997AnGraph}; Figure \ref{arg_review_main_figure_genomearg}D). Considering this representation of an ARG as a set of trees (which we refer to as the \textit{tree representation}) is worthwhile because ARGs are often formulated (see Box \ref{box_tree_sequence}) and operationalized in inference \citep[e.g.,][]{Stern2019AnData,Hejase2022AGraph} based on this representation.  In this tree representation, each non-recombining region has its own local tree that represents the region's evolutionary history. If each recombination breakpoint occurs at a unique position in the genome, as you shift from one local tree to the next (amounting to traversing one recombination breakpoint), the structure of the new tree is identical to its neighbor except for a single edge that is removed and then affixed to a (potentially new) node (Figure \ref{arg_review_main_figure_genomearg}D). In computational parlance, this action is called a \textit{subtree-prune-and-regraft} operation \citep{Song2003OnTrees}. When all recombination events occur at unique locations and each event involves only one breakpoint, the total number of local trees will equal one more than the number of recombination events defining the evolutionary relationships in the genome. For example, in Figure \ref{arg_review_main_figure_genomearg}, two recombination events generate three trees. If recombination events occur at the same location (a breakpoint represents $>$1 recombination event), then moving between adjacent trees will involve a corresponding number of subtree-tree-prune-and-regraft operations (one representing each recombination event), and the tree count will be less than one plus the number of recombination events. 

With inclusion of all nodes involved in recombination and coalescence relevant to the sample nodes, it is straightforward to switch between the two representations. As previously discussed, the local tree for a particular non-recombining region can be extracted from the graphical representation of an ARG by starting at each sample node and tracing the lineages that transmit the region through the ARG until all lineages meet in the MRCA. Conversely, you can recover the graphical representation of an ARG from the local trees by starting with the tree at one end of the set and then sequentially working across the trees, combining the shared nodes and edges, adding the nodes and edges that are not yet included in the graphical structure, and annotating each edge with the non-recombining region(s) that it transmits. As a brief illustration, in Figure \ref{arg_review_main_figure_genomearg}D, the first two trees both contain nodes \circled{\footnotesize S} and \circled{\footnotesize Q} with a connecting edge. In the graphical representation, these shared components would be merged and the edge would be annotated with the transmission of the regions between positions 0 and 2 (as shown in Figure \ref{arg_review_main_figure_genomearg}B).

A recombination event can have several consequences for the structure of adjacent trees. First, it could alter the topology (i.e., the specific branching structure) if the new edge joins to a node on a different edge (e.g., the first and second trees in Figure \ref{arg_review_main_figure_genomearg}D). However, if the new edge joins to a different node on the same edge, the topology will remain unchanged, and only the edge lengths (i.e., coalescent times) will be modified (e.g., the second and third trees in Figure \ref{arg_review_main_figure_genomearg}D). It is also possible for the lineage to coalesce back into the same node, which would result in no change to the tree structure. Each local tree contains every sample node because all samples possess the entire genome (and thus every non-recombining region represented by each tree). However, the collection of nonsample nodes can differ across trees. If an ARG includes all nodes (i.e., every nonsample node is retained), the absence of a node in a local tree signals that it does not represent a genetic ancestor for that region. If an ARG has been simplified (unary nodes removed), the absence of a node either means that it is not a genetic ancestor or that the node does not represent a genome in which coalescence occurred that involved the sample nodes.

There are several key characteristics of an ARG's tree representation. First, the subtree-prune-and-regraft operations that differentiate adjacent trees highlights that nearby trees are generally quite similar and frequently share many nodes and edges \citep{Hudson1991GeneProcess,Rosenberg2002GenealogicalPolymorphisms}. A series of shared nodes and edges between trees indicates that the corresponding non-recombining regions were found in the same lineage in that portion of the ARG. The correlated nature of the trees can be exploited for highly efficient tree storage and computation (\citealt{Kelleher2016EfficientSizes,Kelleher2018EfficientSimulation}; Figure \ref{arg_mutation_fig}A,B; see Box \ref{box_tree_sequence} for further details). Second, although local trees can overlap in structure, a tree can contain components that are not universally found across the entire set of trees (e.g., in Figure \ref{arg_review_main_figure_genomearg}D, node \circled{\footnotesize S} in the first tree is not found in the third tree). The different histories of inheritance mean that each non-recombining region may coalesce in different ancestors that potentially existed at different times in the past and that differ from the GMRCA. For example, in Figure \ref{arg_review_main_figure_genomearg}D, node \circled{\footnotesize X} is the MRCA of the first two trees (the same node as the GMRCA) while the third tree's MRCA is node \circled{\footnotesize W}. If the local trees' MRCAs existed at different times in the past, this will manifest as variation in tree height \citep{Hudson1991GeneProcess}.

Although the information contained in the graphical and tree representations of an ARG is the same, many readers, especially those with a background in phylogenetics, may prefer to think about ARGs via their tree representations. Unlike the graphical representation, each local tree is a familiar object: it is strictly bi- or multi-furcating, meaning that each node has exactly one ancestor and two or more descendants, and that therefore the tree contains no loops (i.e., it is non-reticulate), and is the desired result of a phylogenetic analysis run on a multiple sequence alignment of the DNA in the tree's non-recombining region. Building off this intuition, a phylogeneticist may draw on experience and imagine the set of local trees as analogous to a Bayesian posterior distribution of phylogenies. However, although this intuition may be initially useful, it is important to remember that each local tree is not independent of the others, both because each is generally separated from its neighbors by a small number of recombination events (so is therefore highly correlated), and because the same nodes and edges may appear across multiple local trees. The shared structure of trees imbues the nodes and edges with different properties relative to the analogous components in a standard phylogeny. For example, in a standard phylogeny, branches depict ancestor-descendant relationships through time and thus are one-dimensional. In contrast, edges in an ARG exist both through time and across the genome, and thus can be conceptualized as two-dimensional \citep{Shipilina2023OnBlocks}. This two-dimensionality can be seen in Figure \ref{arg_review_main_figure_genomearg}B where edges extend along the vertical, time dimension and also along different extents of the genome (edges contain different sets of genomic regions). Equivalently, the genome dimension of edges manifests in an ARG's tree representation (Figure \ref{arg_review_main_figure_genomearg}D) through edges persisting across different sets of local trees. The overlapping nature of local trees (i.e., shared nodes and edges) underlies much of an ARG's utility and facilitates the power of ARG-based inference, which we discuss later in the review.

\subsection{Modeling coalescence with recombination}
In population genetics, ARGs are commonly generated by simulating under \citeauthor{Hudson1983PropertiesRecombination}'s (\citeyear{Hudson1983PropertiesRecombination}) model of coalescent with recombination, which is closely connected to the original conception of ARGs \citep{Griffiths1997AnGraph}. Under this model, a set of genomes exists in the present and the lineages describing each genome's ancestry are traced backward in time. Either coalescence or recombination can occur, which represent competing events with exponentially distributed waiting times. With coalescence, two lineages find common ancestry and merge into one. With recombination, a genomic position is selected uniformly as the breakpoint location. The offspring chromosome is inherited from one parental chromosome on one side of the breakpoint and the other parental chromosome on the other side. Recombination splits a lineage into two backwards in time. This process produces a series of genealogies across the genome that describes the ancestry of each genomic position. One question that may arise here is whether recombination could preclude the lineages from finding common ancestry because it increases the lineage count. However, backwards in time, the lineage count grows via recombination at a linear rate ($kR/2$ where $k$ = lineage count and $R$ = recombination rate) whereas lineages coalesce at a quadratic rate [$k(k -1)/2$], and thus finding common ancestry is guaranteed \citep{Griffiths1997AnGraph}. Later in the review, we will be simulating under this model to explore various features of ARGs.

\subsection{ARGs in practice}
In our introduction of ARGs, we mainly focus on the ancestors that are involved in coalescence and recombination. However, when navigating the literature, it is important to recognize that the term \textit{ancestral recombination graph} is frequently applied to structures that differ in various ways from each other and potentially from how we describe ARGs here. This variation stems from both terminological imprecision and inferential limitations.

The degree of completeness in which genetic inheritance from ancestors to descendants is documented can vary extensively. At the most comprehensive extreme, one could record all the genomic material that is passed between ancestors and descendants regardless of whether the material is ancestral or non-ancestral to the samples. Alternatively, one could render an ARG comprehensive to only the focal samples by only keeping track of the material that is ancestral to them (sometimes referred to as a \textit{full ARG}). This structure could be further simplified in various ways such as removing nodes that are unary in one or more local trees. Although these descriptions of ancestry vary in the information that they include, they have all been referred to as ARGs in the literature (Wong et al., unpublished).

Although ARGs may fully document genetic ancestry in theory, we rarely work with such a comprehensive structure in practice. First, in empirical settings, it is not possible to infer all of this information. The sample space of possible structures for a comprehensive ARG quickly becomes impractically vast with increasing genome and sample sizes. Hence, assumptions and shortcuts [e.g., the sequentially Markovian coalescent \citep[SMC;][]{McVean2005ApproximatingRecombination}] are often employed \citep{Rasmussen2014Genome-WideGraphs}, which sacrifices a capacity to infer a comprehensive and fully accurate ARG for the sake of computational tractability. There are also many components of ARGs that are largely unidentifiable and thus are necessarily omitted. For instance, contemporary samples can provide only limited information on unary nodes, and certain features may be imperceptible in contemporary samples. An example of this is a ``diamond'' structure \citep{Rasmussen2014Genome-WideGraphs}, where (going backward in time) recombination splits a lineage but then the lineages immediately coalesce again. Additionally, many sites in the genome are uninformative regarding the local tree topologies (e.g., invariant and singleton sites), which frequently precludes the identification of precise recombination breakpoint locations and other ARG features. More generally, patterns of shared variants represent the information from which ARGs are inferred, while recombination reduces the informative sites per genealogy by dividing the genome into smaller regions. ARG inference will therefore tend to decline in accuracy when the ratio of mutations to recombination is low \citep{Hubisz2020InferenceARGweaver}. This tension between mutation and recombination imposes a theoretical limit on ARG recoverability from sequencing data \citep{Hayman2023RecoverabilityTopologies}.

As a consequence of these obstacles, in practice, we are restricted in what we can infer about genetic ancestry from genomic data. For example, \texttt{tsinfer} \citep{Kelleher2019} infers the collection of local trees and their shared structure (i.e., how nodes and edges overlap across trees) by first estimating ancestral haplotypes and then deducing the tree topologies by inferring how haplotypes relate to each other. This output can be thought of as representing the outcome of coalescence and recombination rather than completely encoding the events that generated the relationships \citep{Kelleher2019}. That is, we are inferring the relationships across the genome produced by recombination and coalescence, but we lack detail on the recombination events that determine how these genealogies exactly knit together in an ARG. Importantly, even if we can acquire comprehensive information on genetic ancestry (e.g., in a simulation), many questions may only require certain subsets of this information, such as the structure of local trees. To accommodate both the existing terminological ambiguity and the realities of how well we can infer genetic ancestry, we permissively apply the term \textit{ancestral recombination graph} to encompass structures that document genetic ancestry in the presence of recombination at varying levels of completeness.

\section{Deepening ARG intuition with simulations}
\label{sec:arg_sim}
To further develop a foundational intuition for ARGs and reinforce content covered in the primer section, we implemented a series of simulations in \texttt{msprime} v1.2.0 \citep{Baumdicker2022Efficient1.0} using the classical coalescent with recombination model. We completed post-simulation processing, analysis, and visualization using \texttt{tskit} \citep{Kelleher2018EfficientSimulation}, \texttt{numpy} \citep{Harris2020ArrayNumPy}, and \texttt{pandas} \citep{McKinney2010DataPython} in \texttt{Python} 3.11.2 \citep{PythonSoftwareFoundation2023Python} and the following packages in \texttt{R} 4.2.3 \citep{RCoreTeam2023R:Computing}: \texttt{TreeDist} \citep{Smith2023TreeDist:Treesb}, \texttt{ape} \citep{ape}, \texttt{ggtree} \citep{Yu2017Ggtree:Data}, \texttt{dplyr} \citep{dplyr}, \texttt{ggplot2} \citep{Wickham2023Ggplot2:Graphics}, \texttt{ggforce} \citep{Pedersen2022Ggforce:Ggplot2}, and \texttt{ggridges} \citep{Wilke2022Ggridges:Ggplot2}. We include all code in the paper’s associated repository and on github (\url{https://github.com/AlexLewanski/arg_review}).

First, to illustrate several general features of ARGs, we focus on a single simulation involving one population with an effective population size of 100 diploid individuals, a genome size of 10 kilobases (kb), a sample size of 10 diploid individuals, and a uniform recombination rate of \num{5e-5} per base per generation. In the simulation, we recorded the full ARG, in which all nodes involved in common ancestry and recombination are retained. We then simplified the ARG structure, which involves removing unary nodes so that remaining nodes represent those that correspond to at least one coalescence event in the genome. Across the 593 local trees generated from this simulation, tree height (TMRCA of each non-recombining region) varied between 57.29 and 1,214.71 generations (non-integer generations are possible here because simulations involved a continuous time model) with a mean$\pm$standard deviation of 448.87$\pm$209.38 generations. The step-like pattern of tree height along the genome, in which height is constant for a stretch, then suddenly jumps to another value, appears because each tree (with a single height) applies to all sites in each non-recombining region (Figure \ref{arg_intuition_sim_fig}A). As discussed in the primer section, another ubiquitous feature of the ARG is that nearby local trees are often highly similar. As a simple illustration of this, we quantified the dissimilarity of all pairwise combinations of local trees using the (approximate) subtree-prune-and-regraft (SPR) distance \citep{Hein1996OnTrees,deOliveiraMartins2008PhylogeneticTrees}, which is the minimum number of subtree moves required to convert one tree to another only based on tip identities (ignoring identities of internal nodes). The topologies of nearby trees were highly similar, with similarity rapidly attenuating with increasing breakpoint separation between trees (Figure \ref{arg_intuition_sim_fig}B). This can also be seen in the matrix of SPR distance values (Figure \ref{arg_intuition_sim_fig}C), with lower values clustered around the diagonal (trees with similar indices and few intervening non-recombining regions) and values rapidly increasing away from this region. The attenuating similarity can also be qualitatively observed in the example trees included in Figure \ref{arg_intuition_sim_fig}A, where the second and third trees, which are adjacent (the 437th and 438th trees, respectively), appear highly similar and are both clearly different in structure compared to the more distant first (45th) and fourth (576th) trees.

\begin{figure}[btp]
    \captionsetup{labelformat=one-page}
    \centering
    \includegraphics[width=\textwidth]{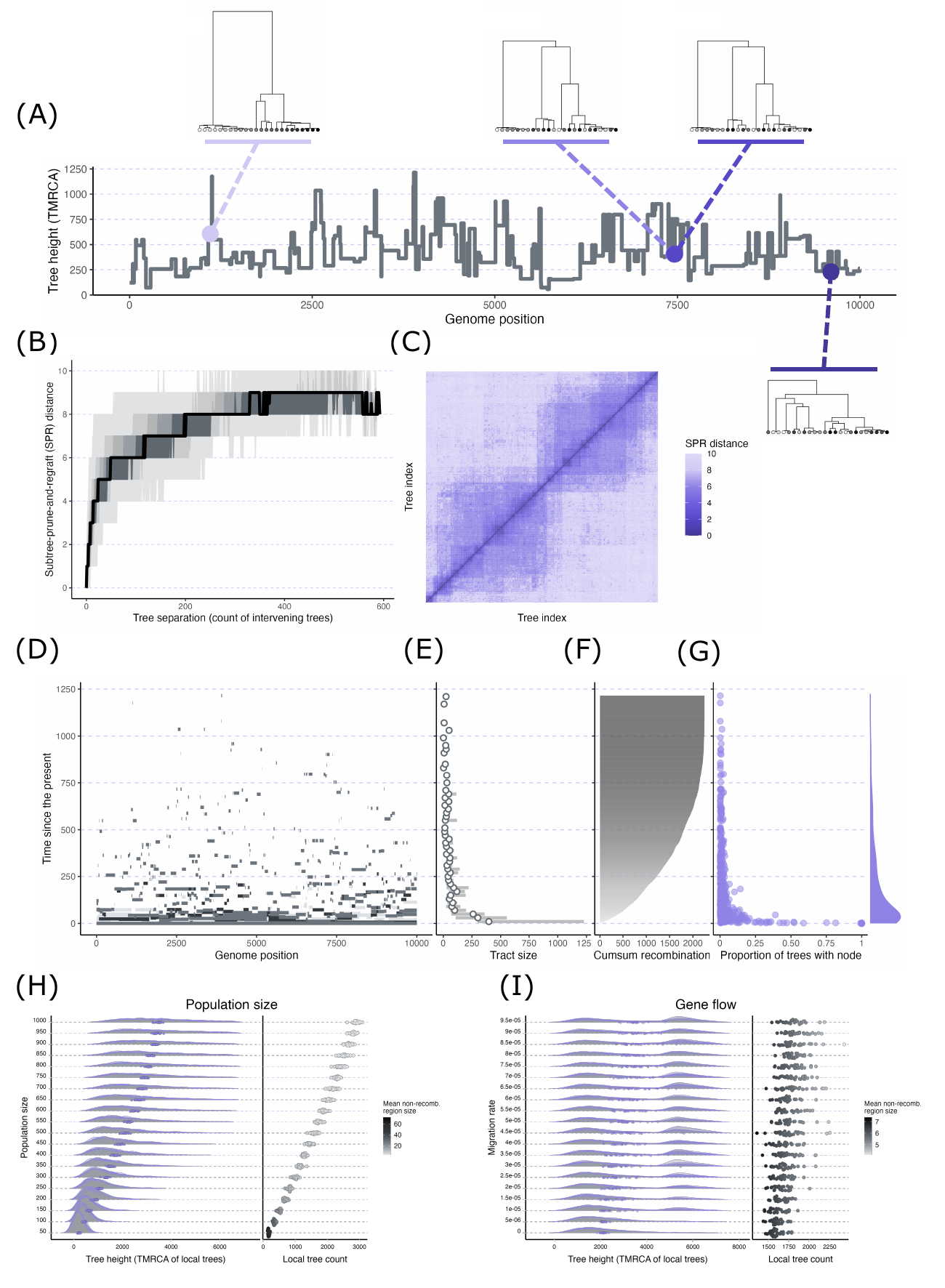}
    \caption{}
\end{figure}
\begin{figure}
    \captionsetup{labelformat=two-page}
    \ContinuedFloat
    \caption{Exploration of ARGs via coalescent simulations. Panels (A)--(C) visualize summaries for a single population simulation. (A) Plot of tree height (TMRCA) along the genome with several example trees plotted along this sequence. (B) The topological dissimilarity of all pairwise combinations of trees was quantified with subtree-prune-and-regraft (SPR) distance. The plot shows SPR distance vs. the number of non-recombining regions separating each tree. The different shaded bands correspond to different percentiles of SPR distance values at each tree separation count: 0--100 (lightest gray), 10--90, 20--80, 30--70, 40--60, 50 (black line). (C) Matrix of SPR distances for all combinations of trees organized by tree index (e.g., the 30th tree in the genome has an index of 30). (B) and (C) illustrate how nearby local trees are highly similar with similarity rapidly declining with growing number of breakpoints separating the trees. (D) Tracking the genomic material for three sample nodes back in time through their genetic ancestors (each node's ancestral material is shown in a different shade of gray). Continuous tracts of ancestral material get progressively smaller back in time as recombination repeatedly breaks the tracts into smaller pieces. (E) The size of tracts of ancestral material swiftly declines going back in time. The plot shows the mean (points) and 25th/75th percentiles of tract size (gray bars) for 20 generation bins. (F) The cumulative number of recombination events occurring backwards in time. (G) The number of nodes and node sharing across local trees in an ARG quickly decline backward in time. The plot shows the location of each node in time (vertical axis) versus the proportion of local trees that contains each node (horizontal axis). The marginal density plot along the vertical axis shows the distribution of nodes through time. (H) A series of simulations with all conditions held constant except for population size. (I) A series of simulations with all conditions held constant except for gene flow rate. The left plots in (H) and (I) show the distribution of tree height for each population size or migration value with the purple points representing the mean value per single simulation run. The right plots in (H) and (I) show the mean tree count per simulation run with each point shaded with its mean non-recombining region size.}
    \label{arg_intuition_sim_fig}
\end{figure}

Next, using the same simulation, we tracked genetic material found in the contemporary sample nodes (hereafter \textit{ancestral material}) back in time through the samples' ancestors. Because we simplified the ARG, tracts of ancestral material identified for a particular sample node also represent tracts of common ancestry (i.e., the material is ancestral to at least one other sample node). For three sample nodes, Figure \ref{arg_intuition_sim_fig}D displays the location of ancestral material (horizontal axis) and the timing of the ancestors carrying that material (vertical axis). At the contemporary time point (time = 0), the tracts of ancestral material span the entire genome because these represent the sample nodes that by definition possess their entire genome as a single haplotype. Traveling back in time (up the vertical axis in Figure \ref{arg_intuition_sim_fig}D), the tracts of ancestral material are broken up into small pieces. Consequently, the average tract length of ancestral material peaks in the contemporary time period and rapidly declines back in time (Figure \ref{arg_intuition_sim_fig}E). This pattern emerges because the cumulative number of recombination events that have occurred in the transmission of ancestral material grows through time (Figure \ref{arg_intuition_sim_fig}F), resulting in the fragmentation of ancestral material into progressively smaller pieces. 

This pattern can also be understood through the lens of node-sharing across the local trees. At the present, every node is shared across all trees because all regions of the genome are found in each sample node. However, moving back in time, the tracts of ancestral material become progressively smaller and thus span fewer non-recombining regions. This results in a decline in node-sharing across trees further back in time; any particular node is carrying ancestral material for a decreasing number of non-recombining regions. Figure \ref{arg_intuition_sim_fig}G depicts this phenomenon. Nodes with the highest proportion of sharing between trees are exclusively located near the present, while nodes located further back in time (higher up the vertical axis) show low proportions of sharing. The reduced node-sharing through time corresponds to variation in how quickly the trees change at different time periods. Near the present, the high degree of node-sharing means that tree structures remain fairly stable. However, the more rapid turnover of nodes at deeper time points translates into faster changes as you move across the trees further back in time. 

A variety of variables can systematically modify features of an ARG. As a brief illustration, we examined how effective population size and gene flow, which frequently vary across studies and systems, influence three fundamental features: tree height, the number of local trees, and the size of non-recombining regions in an ARG. For the population size demonstration, we completed a set of simulations that kept all variables constant (sequence length = 10 kb, recombination rate = \num{3e-5} per base per generation, sample size = 10 diploid individuals) except for population size, which varied between 50 and 1,000 in increments of 50 (a total of 20 population sizes with 30 replicates per size). Tree height and local tree count both increased while mean region size decreased at greater population sizes (Figure \ref{arg_intuition_sim_fig}H). The correlations between population size and the three variables emerge because, with higher effective population sizes, coalescent times will tend to increase \citep{Coop2020GeneticDiversity} because more individuals exist that act as possible ancestors and thus there is a lower probability of any two lineages finding common ancestry in a particular generation. Because of the deeper coalescent times (which result in greater tree heights), more opportunities exist for recombination to occur, which results on average in more local trees and smaller non-recombining regions.

We generated another set of simulations for the gene flow demonstration where we kept all variables constant (sequence length = 10 kb, recombination rate = \num{3e-5} per base per generation) except for migration. We simulated two populations of 500 individuals each that merged (backwards in time) after 5,000 generations. While the populations were separated, one of the populations (the \textit{recipient population}) experienced continuous, unidirectional gene flow from the second population (the \textit{donor population}) forward in time. We varied the migration rate between 0 and \num{1e-4} in increments of \num{5e-6} (a total of 20 different migration rates with 30 replicates per rate). We then sampled 10 diploid individuals from the recipient population. With increasing gene flow, trees tended to increase in height on average, which was associated with increasing bimodality in the distribution of tree heights. This bimodality phenomenon emerges because the presence of two populations along with gene flow result in two distinct time periods during which lineages can coalesce \citep{Maruyama1970EffectivePopulation,Rosenberg2002TheTimes}. The left mode of the distribution corresponds to non-recombining regions whose entire history postdating the population split occurred within the recipient population, and thus coalescence for that region could occur fairly rapidly (small TMRCA values). However, with gene flow, part of a region's history can occur in the donor population. Consequently, a region whose ancestry involves the donor population must wait until the two populations merge in the ancestral population before finding its MRCA. This results in the second, later mode in tree heights. The slight trends of increasing tree count and decreasing region size at greater migration rates occur because the tree heights are increasing on average, which provides opportunities for more recombination events.

Note that the ARG summaries we have reported here---tree height, number of local trees, length of non-recombining regions, similarity and node-sharing between local trees---only represent a small glimpse into the innumerable ways that ARGs can be dissected and summarized. We chose this set to exemplify fundamental features of ARGs and illustrate how they reflect and can therefore be informative about demographic and evolutionary phenomena that are frequently of interest in evolutionary genomics.

\section{ARGs in evolutionary genomics}
From a practical perspective, two questions logically ensue from the ARG introduction: what is the utility of ARGs in evolutionary genomics, and what advantages does it impart relative to existing approaches? As with many methodological advances, ARGs can offer multiple benefits, including strengthening our ability to answer existing questions and opening up entirely new fields of inquiry.

\begin{figure}[btp]
    \captionsetup{labelformat=one-page}
    \centering
    \includegraphics[width=\textwidth]{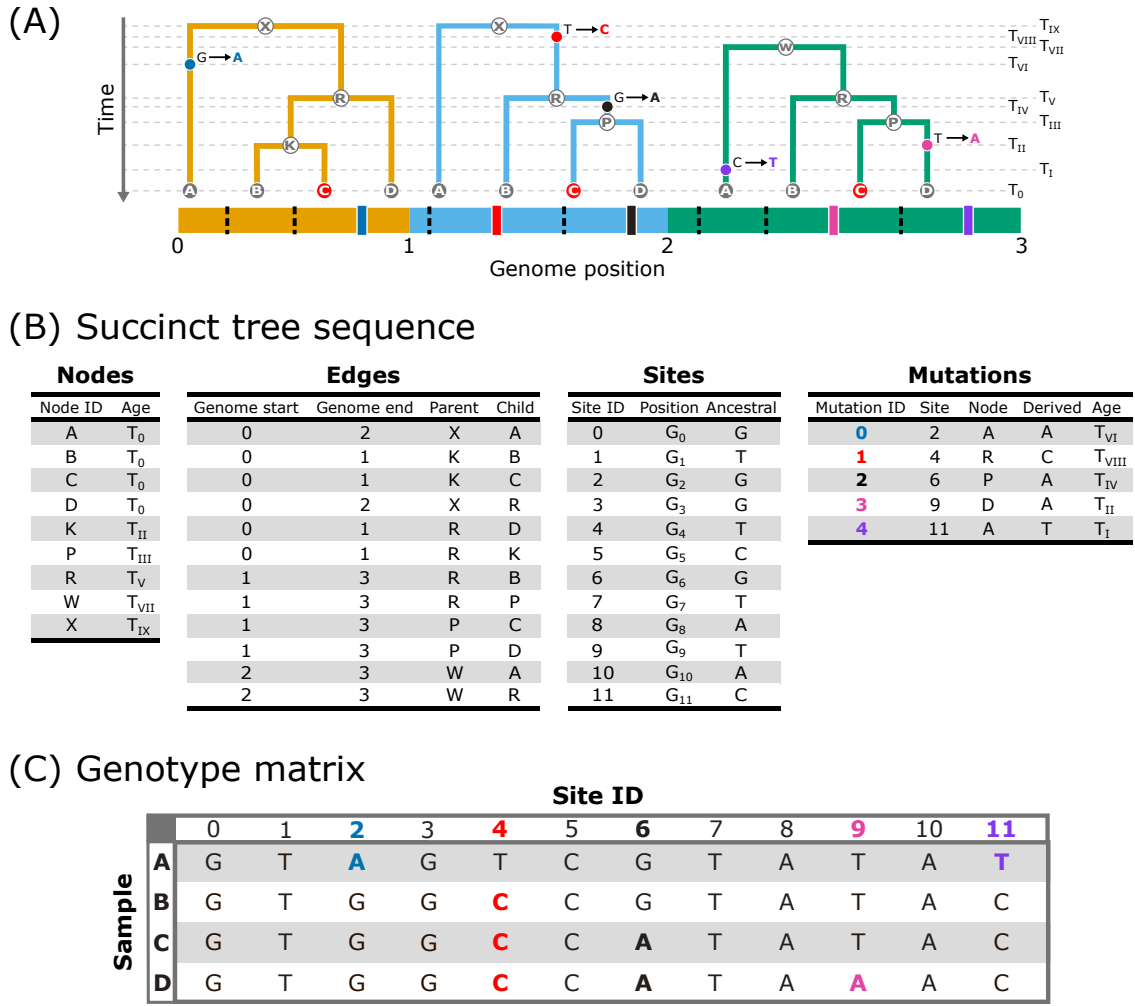}
    \caption{The encoding of local trees and genotype data in the succinct tree sequence format. (A) Depiction of the local trees shown in Figure \ref{arg_review_main_figure_genomearg} with timing and location of mutation events mapped onto the branches and the location of each site shown on the genome. The black, dashed lines represent the invariant sites and the thicker, solid lines represent variant sites corresponding to each mutation. The trees are annotated with horizontal, dashed lines (labelled $T_{0}$--$T_{IX}$) that denote either the timing of coalescence or mutation events. (B) The trees and genotype data in the succinct tree sequence format. The trees are specified with the nodes and edges tables. The nodes table contains an ID and age for each node. The edges table contains the left (\textit{Genome start}) and right (\textit{Genome end}) positions of the genome over which each edge persists, while the \textit{Parent} column contains the nodes that transmit material to the nodes in the \textit{Child} column. The genotypic information is included in the sites [genomic position of each site (\textit{Position}), ancestral state (\textit{Ancestral})] and mutations [derived state (\textit{Derived}), mutation timing (\textit{Age})] tables. (C) The equivalent genotype data for the four sample nodes stored in a more conventional matrix format with the rows representing each sample node and the columns representing each genomic site. Note that with small amounts of genetic data such as this simple example, the tree sequence may require more storage space than a standard genotype matrix format. However, when considering realistic genomes, the tree sequence rapidly becomes much more efficient at storing genetic data with growing sample sizes \citep{Kelleher2019}.}
    \label{arg_mutation_fig}
\end{figure}

To understand how ARGs facilitate empirical inferences that are equal or superior to existing approaches, it is helpful to consider two topics: (1) how ARGs are shaped by evolutionary phenomena, and (2) how ARGs juxtapose with the paradigm of inquiry that currently predominates evolutionary genomics. A critical idea is that the genealogies underlying the genome are the ultimate record of evolutionary history. The structure of an ARG is governed by processes, including selection, drift, and gene flow, that regulate the fitness and relatedness of haplotypes. The genomic composition of individuals is precisely reflected in an ARG’s structure because ARGs encode the ancestral source(s) of samples’ genomes, including how new mutations are propagated through time and across individuals (Figure \ref{arg_mutation_fig}A). Consequently, the genomes of sampled individuals and any summary of their content represent derivatives of the underlying ARG, and many of these genomic summaries can be reinterpreted as explicit descriptions of the ARG \citep{Ralph2019AnData,Ralph2020EfficientlyGenomes}.

Currently in evolutionary genomics, genomic data are typically stored as a genotype matrix [e.g., a VCF file \citep{Danecek2011}; Figure \ref{arg_mutation_fig}C]. The data are distilled down to a variety of summaries such as principal components \citep{Menozzi1978SyntheticEuropeans,McVean2009}, $F$-statistics \citep{Reich2009,Patterson2012,Peter2016AdmixtureF-statistics}, or the site frequency spectrum (SFS) that each reflect particular attributes of the samples' genomes. From these measures, we attempt to infer past phenomena (e.g., selection, demographic changes) that gave rise to the observed data, under the premise that disparities in the generative process translate to corresponding differences in genomic summaries. Indeed, these summaries can often provide powerful and accurate insights into evolutionary processes, and the field of statistical population genetics has made extraordinary strides in divining evolutionary processes from summaries of genetic and genomic data in the six decades since the first empirical measurements of molecular genetic variation were made \citep{Hubby1966APSEUDOOBSCURA}. As previously discussed, each summary measure calculated from these data (e.g., the SFS, $F_{ST}$, $\pi$, $\theta$, individual heterozygosity, identity-by-state, identity-by-descent, etc.) is a low-dimensional summary of an ARG, so, to the extent that we are able to accurately infer an ARG \citep{Brandt2022EvaluationGraphs}, we can recover any of these quantities at least as accurately as they are estimated from the genomic data from which an ARG is inferred \citep{Ralph2020EfficientlyGenomes}. [See \cite{Ralph2019AnData} and \cite{Ralph2020EfficientlyGenomes} for instructive discussions of the ways common summaries of genomic data (and many other quantities) can be calculated and interpreted with ARGs.] And, because ARGs can offer computational efficiencies over traditional methods of storing genomic data, in many cases these quantities can be calculated more easily, and with less computational overhead, from ARGs \citep{Ralph2020EfficientlyGenomes,Nowbandegani2023ExtremelyStudies}.

In some cases, summaries of genomic data made from ARGs can outperform those made from the data directly. For instance, \cite{Nowbandegani2023ExtremelyStudies} devised a method to efficiently represent linkage disequilibrium (LD) based on genomic genealogies (\textit{LD graphical models}). These LD graphical models enable orders-of-magnitude reductions in computation time and memory usage for LD matrix computations and facilitate better polygenic prediction compared to a similar method using the LD correlation matrix. As another example, \cite{Link2023Tree-basedMatrices} found that an expected genetic relatedness matrix calculated from an ARG in a given genomic region more accurately captures relationships than the empirical genetic relatedness matrix calculated in the same region. The higher accuracy may seem counterintuitive; after all, empirical ARGs are estimated from genomic data, so how could statistical inferences conducted on an ARG be \textit{more} accurate than those made directly from the genotype matrix? To see how this can be the case, consider the structure of the genealogies that comprise an ARG. Each local tree is usually separated from that of the adjacent non-recombining region by a small number of recombination events, leading to high correlation in the genealogical relationships contained in nearby trees (e.g., Figures \ref{arg_review_main_figure_genomearg}E; \ref{arg_intuition_sim_fig}B,C). Because of this correlation, the other trees contain information about relatedness between samples in a focal tree. The mutational process is intrinsically random, so that the true genealogical relationships between a set of samples may not be apparent in patterns of shared variation associated with any particular region. By leveraging the information about relationships between samples contained across the entire set of trees, we can, in principle, side-step some of the ``noise" in the data that exists due to the randomness of the mutational process \citep{Ralph2020EfficientlyGenomes}.

Beyond facilitating more efficient and accurate population genetic inferences, the increasing availability of empirical ARGs will foster entirely new fields of ARG-based inquiry. A useful analogy here is the way in which the field of phylogenetics opened up the associated field of phylogenetic comparative methods. For example, the question of whether diversification rates vary across a phylogeny \citep{Ricklefs2007,Rabosky2014AutomaticTrees} is impossible to pose, let alone answer, without a phylogeny. It is difficult to guess what form the ``comparative methods" field of ARGs (i.e., not just asking existing questions better or faster, but entirely new questions that are predicated on ARGs) will take, especially as empirical ARG inference is still in its infancy. However, we can highlight one particularly exciting direction that has already begun to materialize: geographic inference with ARGs.

The recent advances in the reconstruction of genomic genealogies have sparked a revolution in spatial population genetics. In particular, several recent approaches \citep{Osmond2021EstimatingGenealogies,Wohns2022AGenomes} have begun to explore the feasibility of inferring the locations of the genetic ancestors of sampled individuals across space and through time. Although similar geographic inference has been done using non-recombining gene regions \citep[e.g.,][]{Neigel1993ApplicationVariation.,Barton1995GenealogiesGeography, Avise2009Phylogeography:Prospect} or a single phylogenetic tree [``phylogeography" \citep{Knowles2009StatisticalPhylogeography}], it is only with an ARG in hand that it has become feasible to infer locations for \textit{all} the genetic ancestors of a sample. This power, in turn, has facilitated massively more detailed and nuanced understanding of how organisms move across space and through time. For example, \cite{Osmond2021EstimatingGenealogies} inferred the mean effective dispersal distance of \textit{Arabidopsis thaliana}, and \cite{Wohns2022AGenomes} recovered the broad strokes of human dispersal history over the last 800,000 years. In the future, this type of inference of ancestral locations could empower specific and biologically principled definitions of ``admixture" (e.g., 12.5\% of the genetic ancestors of a focal individual are estimated to have lived inside a particular geographic region at a particular slice of time) \citep{Bradburd2019SpatialTime}. The exciting enterprise of geographic inference of ancestor locations (more precisely, of the geographic locations of nodes in an ARG) and of the concomitant historical patterns of dispersal and density described by a sample's georeferenced genealogy, is entirely predicated on the existence of an inferred ARG for a set of samples.

An important qualifier to this discussion is that, despite the evident promise of ARG-based inference, it remains less clear the extent to which this promise is achievable in empirical biology. One of the main reasons for this uncertainty is, despite some awareness of empirical limits on ARG reconstruction, little is known regarding the degree of accuracy needed to make quality downstream inferences from ARGs. For example, do accurate inferences generally presuppose highly precise and accurate estimates of ARGs? Or perhaps some questions only require accuracy in specific properties of ARGs. For example, the distribution of local tree TMRCAs may need to be accurate \citep{Brandt2022EvaluationGraphs}, while the accuracy of their topologies are less crucial. Understanding the sensitivities and requirements of downstream inferences will help uncover the particular facet(s) of ARG reconstruction whose improvements would be most beneficial and will also help delineate the limits that empirical ARG reconstruction will enforce on downstream inferences.

\section{Conclusions}
In this review, we aimed to introduce ARGs, articulate the capacity of ARGs to enhance the study of evolutionary genomics, and describe the current and/or forthcoming practicability of using ARGs in empirical- and simulation-based research. Indeed, ARGs have the potential to advance empirical evolutionary genomics in both minor and profound ways ranging from improving implementation of existing approaches (e.g., faster calculation of traditional population genetics statistics) to inspiring novel and previously inaccessible avenues of study. The nature and extent to which ARGs will reshape the field remains unclear and will depend on fundamental limits regarding the information contained in empirical ARGs, the degree to which ARGs are integrated into the methods canon of evolutionary genomics, and our collective ingenuity.

How do we fully capitalize on ARGs? First, a broader suite of inference methods and tools based on ARGs must (continue to) be developed, evaluated, and made readily accessible to the broader community. Until now, most ARG-based methods development has concentrated on ARG reconstruction and simulation. Although these topics will benefit from additional progress, we are reaching a stage where empirical- and simulation-based ARGs can be realistically acquired in many situations and readily stored and manipulated with an increasingly mature and powerful software infrastructure  (e.g., \texttt{tskit}). A more expansive body of methods built on ARGs will enable wider adoption of ARG-based inference. The incipient nature of ARG methods presents an opportunity for more extensive synthesis and synergy between evolutionary genomics and both phylogenetic comparative methods and phylogeography. These fields have developed a sizeable assortment of phylogenetic methods that could be co-opted and modified for tree-based inference in the context of ARGs. Fully capitalizing on our growing ARG capabilities will clearly require a receptivity to new genealogically explicit approaches and ideas that have so far only featured sparingly in empirical evolutionary genomics. However, with a concerted embrace of ARGs, we are confident that this ``holy grail of statistical population genetics'' \citep{Hubisz2020InferenceARGweaver} will further realize its potential for many questions in evolutionary biology.

\clearpage
\newpage
\section{Boxes}

\begin{info_box}[enhanced jigsaw,breakable,pad at break*=1mm,before upper={\parindent15pt},colback=gray!5!white,colframe=gray!75!black,label=box_arg_reconstruction]{ARG reconstruction}


A growing arsenal of methods is available to infer ARGs from genomic data. \texttt{ARGweaver}, which was introduced in \citeyear{Rasmussen2014Genome-WideGraphs} by \citeauthor{Rasmussen2014Genome-WideGraphs}, represents a seminal achievement in ARG inference. \texttt{ARGweaver} and its extension (\texttt{ARGweaver-D}; \citealt{Hubisz2020MappingGraph}) leverage approximations of the coalescent [SMC or SMC' \citep{McVean2005ApproximatingRecombination,Marjoram2006FastSimulation}] and time discretization to simplify the space from which to sample candidate ARGs using Markov Chain Monte Carlo. These methods, along with other recent Bayesian approaches like \texttt{Arbores} \citep{Heine2018BridgingGraphs} and \texttt{ARGinfer} \citep{Mahmoudi2022BayesianGraphs}, enable the rigorous treatment of uncertainty via the incorporation of an ARG's posterior distribution into downstream analyses. One general limitation of these methods is that, due to computational requirements, they can only handle fairly modest sample sizes. For example, \texttt{ARGweaver} can consider between two to about 100 samples \citep{Hubisz2020InferenceARGweaver}.

Motivated by the extensive sequencing efforts in human genomics, several methods have been devised to accommodate large and complicated genomic datasets. For example, \texttt{ARG-Needle} \citep{Zhang2023Biobank-scaleTraits}, \texttt{tsinfer}/\texttt{tsdate} \citep{Kelleher2019,Wohns2022AGenomes}, and \texttt{Relate} \citep{Speidel2019ASamples} can infer genomic genealogies for tens of thousands (\texttt{Relate}) to hundreds of thousands (\texttt{tsinfer}, \texttt{ARG-Needle}) of human samples. \texttt{Relate} and \texttt{tsinfer} can additionally incorporate samples from different time periods and have been used to reconstruct unified genomic genealogies for modern humans and ancient samples of humans, Neanderthals, and Denisovans \citep{Speidel2021InferringGenealogies,Wohns2022AGenomes}. This scalability is facilitated by various statistical simplifications, which result in several limitations in the inferences of these approaches. For example, \texttt{Relate} and \texttt{tsinfer} infer less information about recombination than methods like \texttt{ARGweaver}, which attempts to identify the specific recombination events associated with every breakpoint (Wong et al., unpublished). Additionally, they only provide point estimates for the tree topologies, which precludes comprehensive assessments of uncertainty in ARG structure.

So far, most ARG inference development has focused on human and other eukaryotic genomes. However, there are also active efforts to create methods tailored to other types of genomes and systems. For example, \cite{Vaughan2017InferringData} developed a Bayesian approach dubbed \texttt{Bacter}, which is designed to infer ARGs for bacteria based on the ClonalOrigin model \citep{Didelot2010InferenceSequences}. Spurred by the COVID-19 pandemic, \cite{Zhan2023TowardsSARS-CoV-2} recently introduced a method (\texttt{sc2ts}) for ARG reconstruction that can involve millions or more of SARS-CoV-2 genomes. \texttt{sc2ts} is designed to construct and repeatedly update the ARG through time with new samples, which is relevant to ongoing surveillance during pandemics wherein pathogen samples are collected and sequenced in real time.

In summary, there is a burgeoning assortment of methods that enable ARG reconstruction across a range of dataset and system characteristics including data types, sample sizes, and sampling regimes. ARG reconstruction remains a formidable statistical and computational challenge, and many improvements in the robustness and flexibility of ARG reconstruction are still needed \citep[e.g.,][]{Deng2021TheGraphs,Brandt2022EvaluationGraphs,Ignatieva2023TheGraphs}. However, ARG inference has emerged as a nexus of methodological development in statistical population genetics, and ongoing efforts exist to address the limitations and combine the strengths of existing methods \citep[e.g.,][]{Rasmussen2023Espalier:Forests}. Readers should be prepared for continued innovation in this area.

\end{info_box}

\newpage
\begin{info_box}[enhanced jigsaw,breakable,pad at break*=1mm,before upper={\parindent15pt},colback=gray!5!white,colframe=gray!75!black,label=box_arg_simulation]{Simulation}

Concurrent with improvements in ARG reconstruction, revolutionary progress in population genomic simulation has occurred over the past decade. One of the most significant developments was \texttt{msprime} \citep{Kelleher2016EfficientSizes,Baumdicker2022Efficient1.0}, which can simulate the genomes and ancestry for a set of samples backwards in time using the coalescent. With the coalescent, only the ancestors of the samples (and not entire populations) need to be tracked. This approach is highly efficient but generally entails an assumption of neutral evolution [although it is possible for coalescent theory and simulation to incorporate selection \citep[e.g., ][]{Kaplan1988TheSelection,Hudson1988TheRecombination.,Walczak2012TheCoalescent,Spencer2004SelSim:Recombination,Kern2016Discoal:Selection,Baumdicker2022Efficient1.0}]. The notable innovation of \texttt{msprime} relative to previous coalescent programs is the speed at which it can perform simulations at biologically realistic scales under a variety of models and with recombination. For example, \texttt{msprime} has been used to simulate realistic whole genome sequences based on genealogical information for $\sim$1.4 million people inhabiting Quebec, Canada \citep{Anderson-Trocme2023OnQuebec}.

Another noteworthy development in population genomic simulation over the past decade is \texttt{SLiM} \citep{Messer2013SLiM:Linkage}. In contrast to coalescent simulators, \texttt{SLiM} simulates forward in time using either Wright-Fisher or non-Wright-Fisher models \citep{Haller2019SLiMModel}. The forward-in-time nature of \texttt{SLiM} means that all individuals in each generation (including historical individuals that are not genetic ancestors to the contemporary population) must be tracked in the simulation. This elevates the computational burden compared to coalescent simulation. However, it enables substantially more flexibility in the scenarios that can be simulated including complex selection and ecological interactions across multiple species \citep{Haller2023SLiMModeling}. Relevant to this review, both \texttt{SLiM} and \texttt{msprime} can record ARGs during simulation \citep{Haller2019Tree-sequenceGenomes,Baumdicker2022Efficient1.0}. These and other simulation programs [e.g., \texttt{discoal} \citep{Kern2016Discoal:Selection}] can be used for a variety of purposes in ARG-based research including exploration of biological phenomena, statistical and machine learning inference \citep[e.g.,][]{Hejase2022AGraph,Campagna2022SelectiveSpecies,Korfmann2023SimultaneousCoalescent}, and methods evaluation \citep{Brandt2022EvaluationGraphs}.

\end{info_box}

\newpage

\begin{info_box}[enhanced jigsaw,breakable,pad at break*=1mm,before upper={\parindent15pt},colback=gray!5!white,colframe=gray!75!black,label=box_tree_sequence]{The succinct tree sequence}

The correlated nature of an ARG’s local trees can be exploited to compactly encode the trees in a data structure termed the \textit{succinct tree sequence} or \textit{tree sequence} for short \citep[Figure \ref{arg_mutation_fig}A,B;][]{Kelleher2016EfficientSizes,Kelleher2018EfficientSimulation}. The tree sequence defines the trees using two tables. The node table contains an identifier and the timing of each node (first table in Figure \ref{arg_mutation_fig}B). The edge table documents the edges shared between adjoining trees by recording the parent and offspring nodes of each edge and the contiguous extent of the genome that each edge covers (second table in Figure \ref{arg_mutation_fig}B). The key innovation here is that the data structure eliminates substantial redundancy. Instead of storing each tree independently, which would necessitate duplication of shared nodes and edges, the tree sequence records each shared component just once. 

The basic tree sequence technically does not encode the full ARG, which includes all coalescent and recombination events. The basic tree sequence only explicitly contains information on the coalescent events and does not detail the timing and specific changes that differentiate adjacent trees. \cite{Kelleher2019} explain this distinction as follows: the full ARG ``encodes the events that occurred in the history of a sample'' while the set of local trees recorded in the tree sequence ``encodes the outcome of those events.'' Nonetheless, the tree sequence can be elaborated with recombination information to more exhaustively document genetic ancestry \citep[e.g.,][]{Baumdicker2022Efficient1.0,Mahmoudi2022BayesianGraphs}.

Several properties of the tree sequence have revolutionized ARG-based research. First, its concise nature means that an immensity of genealogical information can be stored in a highly compressed manner. The tree sequence is also a flexible format that can be augmented with additional tables to store other information such as location metadata and DNA data (e.g., third and fourth tables in Figure \ref{arg_mutation_fig}B; Figure \ref{arg_mutation_fig}A). Notably, relative to conventional genotype matrix formats (Figure \ref{arg_mutation_fig}C), DNA data can be represented much more efficiently using the tree sequence. For instance, \cite{Kelleher2019} estimated that the tree sequence format could store genetic variant data for 10 billion haploid human-like chromosomes in $\sim$1 TB, which is many orders of magnitude smaller than the $\sim$25 PB required to store these data in a VCF \citep{Danecek2011}. The efficiency of the tree sequence also permits significant speed-ups in computation (e.g., through the implementation of fast algorithms). These features have enabled advancements in the scale and scope of ARG-based analyses and are increasingly accessible given that the tree sequence underpins a growing ecosystem of methods and software including \texttt{tsinfer} \citep{Kelleher2019}, \texttt{sc2ts} \citep{Zhan2023TowardsSARS-CoV-2}, \texttt{ARGinfer} \citep{Mahmoudi2022BayesianGraphs}, \texttt{msprime} \citep{Baumdicker2022Efficient1.0}, \texttt{SLiM} \citep{Haller2019Tree-sequenceGenomes}, and \texttt{tskit} \citep{Kelleher2018EfficientSimulation} built to infer, simulate, and analyze ARGs. Further details on the tree sequence can be found in the papers introducing and expanding the tree sequence \citep{Kelleher2016EfficientSizes,Kelleher2018EfficientSimulation,Mahmoudi2022BayesianGraphs} and in the documentation of \texttt{tskit} \citep{Kelleher2018EfficientSimulation}. 


\end{info_box}


\clearpage
\newpage

\section{Acknowledgements}
We thank Peter Ralph, Sarah Fitzpatrick, Yan Wong, Jerome Kelleher, and members of the Fitzpatrick and Bradburd labs for helpful discussions and manuscript feedback. This work was supported by a University Distinguished Fellowship from Michigan State University (awarded to ALL), the National Defense Science \& Engineering Graduate (NDSEG) Fellowship from the Department of Defense (awarded to ALL). Research reported in this publication was supported by the National Institute of General Medical Sciences of the National Institutes of Health under Award Number R35GM137919 (awarded to GSB). The content is solely the responsibility of the authors and does not necessarily represent the official views of the NIH.

\bibliographystyle{elsarticle-harv} 
\bibliography{references.bib}





\end{document}